\documentclass[aps,prb,twocolumn,superscriptaddress,showpacs]{revtex4}
\usepackage{graphicx}
\usepackage{amsmath}
\usepackage{amsfonts}
\usepackage{times}

\begin{document}

\title{Quasiparticle tunneling in the Moore-Read fractional quantum Hall State}

\author{Hua Chen}
\affiliation{Zhejiang Institute of Modern Physics, Zhejiang
University, Hangzhou 310027, P.R. China}
\affiliation{Asia Pacific Center for Theoretical Physics,
Pohang, Gyeongbuk 790-784, Korea}

\author{Zi-Xiang Hu}
\affiliation{Asia Pacific Center for Theoretical Physics,
Pohang, Gyeongbuk 790-784, Korea}

\author{Kun Yang}
\affiliation{National High Magnetic Field Laboratory and
Department of Physics, Florida State University, Tallahassee,
Florida 32306, USA}

\author{E. H. Rezayi}
\affiliation{Department of Physics, California State University
Los Angeles, Los Angeles, California 90032, USA}

\author{Xin Wan}
\affiliation{Asia Pacific Center for Theoretical Physics,
Pohang, Gyeongbuk 790-784, Korea}
\affiliation{Department of Physics, Pohang University of Science and
Technology, Pohang, Gyeongbuk 790-784, Korea}
\affiliation{Zhejiang Institute of Modern Physics, Zhejiang
University, Hangzhou 310027, P.R. China}

\date{\today}

\begin{abstract}

In fractional quantum Hall systems, quasiparticles of fractional
charge can tunnel between the edges at a quantum point contact.  Such
tunneling (or backscattering) processes contribute to charge
transport, and provide information on both the charge and statistics
of the quasiparticles involved.  Here we study quasiparticle tunneling
in the Moore-Read state, in which quasiparticles of charge $e/4$
(non-Abelian) and $e/2$ (Abelian) may co-exist and both contribute to
edge transport. On a disk geometry, we calculate the matrix elements
for $e/2$ and $e/4$ quasiholes to tunnel through the bulk of the
Moore-Read state, in an attempt to understand their relative
importance.  We find the tunneling amplitude for charge $e/2$
quasihole is exponentially smaller than that for charge $e/4$
quasihole, and the ratio between them can be (partially) attributed to
their charge difference.  We find that including long-range Coulomb
interaction only has a weak effect on the ratio. We discuss briefly
the relevance of these results to recent tunneling and interferometry
experiments at filling factor $\nu=5/2$.

\end{abstract}

\maketitle

\section{Introduction}

The fractional quantum Hall effect (FQHE) at filling factor $\nu =
5/2$~\cite{willett87,Panw99a,Panw99b,Panw99c,Panw99d,Panw99e,Panw99f,Panw99g,Panw99h,Panw99i}
has attracted strong interest, due to the possibility that it may
support non-Abelian quasiparticles, and their potential application in
topological quantum
computation.~\cite{kitaev,Preskillbook,Freedman,Nayakrmp} Numerical
studies~\cite{morf98,Ed2000,wan06,wan08,peterson,moller,feiguin,storni09}
indicate that the Moore-Read (MR) state~\cite{moore91} or its
particle-hole conjugate state\cite{lee07,levin07} are the most likely
candidates to describe the $\nu=5/2$ quantum Hall liquid. They both
support non-Abelian quasiparticle excitations with fractional charge
$e/4$, in addition to the Abelian quasiparticle excitation with
fractional charge $e/2$ of the Laughlin type.~\cite{moore91,nayak96}

Edge excitations in the FQHE can be described at low energies by a
chiral Luttinger liquid model~\cite{wen}, and quasiparticle tunneling
through barriers or constrictions was originally
considered~\cite{kane92,chamonwen} in the case of the Laughlin
state. Recently the transport properties of the $\nu=5/2$ state
through a point contact have also been considered by a number of
authors.~\cite{pfendley,afeiguin,sdas} Experimentally the
quasiparticle charge of $e/4$ has been measured in the shot
noise~\cite{dolevnature} and temperature dependence of tunneling
conductance.~\cite{RaduScience} The latter also probes the tunneling
exponent, which is related to the Abelian or non-Abelian nature of the
state, although a direct probe on the statistics based on
quasiparticle interference is desired.

The two point contact Fabry-P\'erot interferometer was first proposed
for probing the Abelian statistics~\cite{chamon97} and later
considered for the non-Abelian
statistics.~\cite{fradkin98,dassarma05,stern06,bonderson06,rosenow,bishara08,bonderson06b,fidkowski07,bonderson08,ardonne08}
In this kind of setup, quasiparticles propagating along the edges of
the sample can tunnel from one edge to the other at the constrictions
formed in a gated Hall bar.  Such tunneling processes lead to
interference of the edge current between two different tunneling
trajectories. It has been used in both integer~\cite{yji03,yzhang09}
and fractional quantum Hall regimes in the lowest Landau
level.~\cite{camino05,camino07} Recently, Willett
$et~al.$~\cite{willett08,willett09} implemented such a setup in the
first excited Landau level and attempted to probe the non-Abelian
statistics of the quasiparticles in the case of $\nu=5/2$ from the
interference pattern.

The interference pattern at $\nu=5/2$ state is predicted to exhibit an
even-odd variation~\cite{stern06,bonderson06} depending on the parity of
the number of $e/4$ quasiparticles in the bulk. This would be a direct
indication of their non-Abelian nature.  In their experiments, Willett
$et ~al.$~\cite{willett09} observed oscillations of the longitudinal
resistance while varying the side gate voltage in their
interferometer. At low temperatures they observed apparent
Aharanov-Bohm oscillation periods corresponding to $e/4$ quasiparticle
tunneling for certain gate voltages, and periods corresponding to $e/2$
quasiparticle at other gate voltages. This alternation was argued to
be due to the non-Abelian nature of the $e/4$
quasiparticles,\cite{willett09,bishara09} consistent with earlier
theoretical prediction.\cite{stern06,bonderson06} At higher
temperatures $e/4$ periods disappear while $e/2$ periods
persist.\cite{willett08}

There are two possible origins for the $e/2$ period in the
interference picture: It may come from the interference of $e/2$
quasiparticles, or the intereference of $e/4$ quasiparticles that
traverse two laps around the interferometer.  It is natural to expect
that the tunneling of the $e/4$ quasiparticles is much easier than
that of $e/2$ quasiparticles.  Therefore, the tunneling amplitudes of
$e/4$ quasiparticles should be larger than that of the $e/2$
quasiparticles. On the other hand, $e/2$ quasiparticles, being Abelian
(or Laughlin type), involve the charge sector only and have much
longer coherence length than that of $e/4$ quasiparticles.\cite{wan08}
In fact it was predicted\cite{wan08} that the $e/2$ interference
pattern will dominate once the temperature dependent coherence length
for $e/4$ quaisparticles becomes shorter than the distance between the
two point contacts, in agreement with more recent
experiment.\cite{willett08}

In the present paper, we attempt to shed light on the relative
importance of $e/4$ and $e/2$ quasiparticle tunneling in transport
experiments involving point contacts.  By numerically diagonalizing a
special Hamiltonian with three-body interaction that makes the
Moore-Read state the exact ground state at half-filling, we explicitly
calculate the amplitudes of $e/4$ and $e/2$ quasiparticles tunneling from
one edge to another through the Moore-Read bulk state in disk and
annulus geometries.  We find the (bare) tunneling amplitude for charge
$e/2$ quasiparticles is exponentially smaller than that for charge $e/4$
quasiparticles, and their ratio can be partially (but not completely)
attributed to the charge difference. These results would allow for a
quantitative interpretation of the quasiparticle interference pattern
observed by Willett $et~al.$~\cite{willett08}

The remainder of the paper is organized as follows. In
Sec.~\ref{sec:model}, we describe the microscopic model of the
5/2-filling fractional quantum Hall liquid on a disk and its ground
state wave functions with and without a charge $e/4$ or $e/2$
quasihole in the center. We introduce the tunneling potential for the
quasiholes and outline the scheme of our calculation. We then present
our main results for the case of short-range interaction in
Sec.~\ref{subsec:shortrange}, in which we compare the different
tunneling amplitudes of the charge $e/4$ and $e/2$ quasiholes. We also
map the results from the disk geometry to an experimentally more
relevant annulus geometry and attempt to obtain the leading dependence
of the tunneling amplitudes on system size and inter-edge distance. We
discuss the influence of long-range Coulomb interaction in
Sec.~\ref{subsec:longrange}. In Sec.~\ref{sec:discussion}, we summarize
our results and discuss their relevance to recent interference
measurement in the 5/2 fractional quantum Hall system.

\section{Model and Method}
\label{sec:model}

\begin{figure}
\includegraphics[width=7cm]{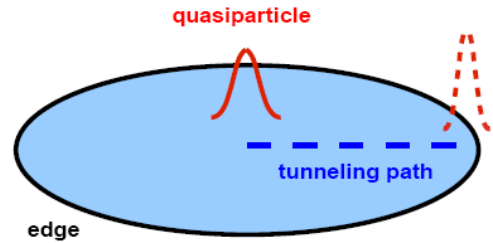}
\caption{\label{fig:setup}(Color online) Theoretical setup for a disk
  of a fractional quantum Hall liquid, allowing quasiholes to tunnel
  through the bulk from the center to the edge.}
\end{figure}

We start by considering a disk on which a $\nu = 1/2$ Moore-Read
fractional quantum Hall liquid resides. The disk geometry can support
both charge $e/4$ and $e/2$ excitations at the center, providing us
with an opportunity to study their tunneling to the edge (see
Fig.~\ref{fig:setup}). Later in the paper, we will also map the
geometry to an annulus or a ribbon of electrons, thus allowing a
closer comparison with realistic experimental situations, e.g. in the
vicinity of a quantum point contact.  Our system resembles a multiply
connected torus with a strong barrier studied
prieviously~\cite{shopen05} in the context of Laughlin quasiparticle
tunneling. In the half-filling case, we need to consider both
Laughlin-type Abelian quasiparticles with characteristic charge $e/2$
and Moore-Read--type non-Abelian quasiparticles with charge $e/4$.

To study the Moore-Read ground states with and without an $e/4$ or
$e/2$ quasihole at a half filling, we start from a three-body interaction
$H_{3B}$
\begin{equation}
\label{eqn:threebody} H_{3B} = -\sum_{i < j <
k}S_{ijk}[\nabla^2_i\nabla^4_j \delta({\bf r}_i - {\bf r}_j)
\delta({\bf r}_i -{\bf r}_k)],
\end{equation}
where $S$ is a symmetrizer:
$S_{123}[f_{123}]=f_{123}+f_{231}+f_{312}$.
The $N$-electron Pfaffian state proposed by
Moore and Read~\cite{moore91} in the lowest Landau level (LLL) representation,
\begin{multline}
\label{eqn:mooreread}
\Psi_{\rm MR} (z_1, z_2, ..., z_N) = \\
{\rm Pf} \left (1 \over {z_i - z_j} \right ) \prod_{i < j} (z_i -
z_j)^2 \exp \left \{ - \sum_i {|z_i|^2 \over 4} \right \},
\end{multline}
is the exact zero-energy ground state of $H_{3B}$ with the
smallest total angular momentum $M_0 = N(2N-3)/2$.
In Eq.~(\ref{eqn:mooreread}), the Pfaffian is defined by
\begin{equation}
{\rm Pf} M_{ij} = {1 \over 2^{N/2} (N/2)!} \sum_{\sigma \in S_N}
{\rm sgn} \sigma \prod_{k=1}^{N/2} M_{\sigma(2k-1)\sigma(2k)}
\end{equation}
for an $N \times N$ antisymmetric matrix with elements $M_{ij}$.

The three-body interaction also generates a series of zero-energy states
with higher total angular momentum, related to edge excitations and
bulk quasihole excitations. The $N$-electron Moore-Read ground state
with an additional charge $e/4$ quasihole at the origin (so the edge
also expand correspondingly due to a fixed number of electrons) has a
wavefunction
\begin{multline}
\label{eqn:eover4}
\Psi_{\rm MR}^{e/4} (z_1, z_2, ..., z_N) = \\
{\rm Pf} \left ({z_i + z_j} \over {z_i - z_j} \right )
\prod_{i < j} (z_i - z_j)^2 \exp
\left \{ - \sum_i {|z_i|^2 \over 4} \right \}.
\end{multline}
This state is a zero-energy state with total angular momentum $M_0 +
N/2$ in the lowest $2N-1$ orbitals (one more than needed for the
Moore-Read state), but not the only one.  To generate the unique
charge $e/4$ state, we need to introduce a strong repulsive
interaction for electrons occupying the lowest two orbitals
\begin{equation}
\label{eqn:repulsion}
  \Delta H_{e/4} =
  \lambda c_1^{+}c_1 c_0^{+} c_0, \,\,\, \lambda \rightarrow \infty.
\end{equation}
On the
other hand, the Moore-Read ground state with a $e/2$ quasihole (i.e.,
a Laughlin quasihole, equivalent to two $e/4$ quasiholes fused in the
identity channel) at the origin,
\begin{equation}
\label{eqn:eover2}
\Psi_{\rm MR}^{e/2} (z_1, z_2, ..., z_N) = \left (\prod_i z_i \right )
\Psi_{\rm MR} (z_1, z_2, ..., z_N),
\end{equation}
is the unique zero-energy ground state
with total angular momentum $M_0 + N$ in the lowest $2N-1$ orbitals.

The Moore-Read state [Eq.~(\ref{eqn:mooreread})], together with its
quasihole states [Eqs.~(\ref{eqn:eover4}) and (\ref{eqn:eover2})], can
therefore be generated by numerically diagonalizing the three-body
Hamiltonian [Eq.~(\ref{eqn:threebody}) with Eq.~(\ref{eqn:repulsion}),
the artificial repulsion to generate an $e/4$ quasihole] in
corresponding finite number of orbitals using the Lanczos
algorithm. The wavefunctions can then be supplied to calculate the
tunneling amplitudes of the quasiholes. The same numerical procedure
can be used to study the tunneling amplitudes for the more realistic
situation with a long-range interaction, in which case the variational
wavefunctions are no longer eigenstates of the realistic
Hamiltonian. For clarity and convenience, we will delay the discussion
on how to generate realistic ground state and quasihole states in the
presence of long-range interaction (and their comparison with the
variational states) to Sec.~\ref{subsec:longrange}.

To study the tunneling amplitudes of the quasiholes, let us first
consider a single-particle picture, which will help us understand our
approach and, later, our results as well.  In the disk geometry, the
single-particle eigenstates are
\begin{equation}
\vert m \rangle \equiv \phi_m(z) = (2\pi2^m m!)^{-1/2} z^m e^{-|z|^2/4}.
\end{equation}
We assume a single-particle tunneling potential
\begin{equation}
V_{\rm tunnel}(\theta) = V_t \delta(\theta),
\end{equation}
which breaks the rotational symmetry.  Here we calculate the matrix
element of $\langle k \vert V_{\rm tunnel}(\theta) \vert l \rangle$,
related to the tunneling of an electron from state $\vert l \rangle$
to state $\vert k \rangle$. One can visualize the tunneling process as
a path along the polar angle $\theta = 0$ between the two states
centered around their maximum amplitude at $|z| = \sqrt{2l}$ and $|z|
= \sqrt{2k}$, respectively. One readily obtains
\begin{equation}
\label{eq:vp0ll}
v_p (k,l) \equiv \langle k \vert V_{\rm tunnel}(\theta) \vert l \rangle
= {V_t \over 2 \pi}
{\Gamma \left ( {{k + l} \over 2} + 1 \right ) \over \sqrt{k! l!}}.
\end{equation}
The interesting limit is that we let $k$ and $l$ tend to infinity, but
keep the tunneling distance fixed at $d$, i.e., $\vert k - l \vert
\sim \sqrt{2k} (d/l_B) \ll (k+l)$.  Alternatively, we can understand
$d$ through the angular momentum change $l_B^2 \vert k - l \vert / R$,
where $R \sim \sqrt{2k} l_B$ is the azimuthal size of the
single-particle state with momentum $k$ (or $l$ in this limit).
We can show (see Appendix~\ref{larged}), in this limit,
\begin{equation}
\label{eq:asymptotic}
  v_p (k,l) \sim {V_t \over 2 \pi} e^{- (k-l)^2 / 4(k+l+2)}
  \sim {V_t \over 2 \pi} e^{-d^2 / (2l_B)^2},
\end{equation}
which reflects the overlap of the two Gaussians separated by a distance
$d$.

For quasiparticle tunneling at filling fraction $\nu = 5/2$, one
should, in principle, use wavefunctions in the first excited Landau
level (1LL). Evaluating the tunneling matrix element in the 1LL, we
obtain an additional prefactor, so
\begin{equation}
\label{eq:vp1ll}
v_p^{1LL} (k,l) = \left [1 - {(k-l)^2 \over 2(k+l)} \right ] v_p (k,l).
\end{equation}
The sign change in the prefactor at $d \sim l_B$ can, unfortunately,
cause severe finite-size effect for the numerically accessible range.
Nevertheless, in the thermodynamic limit, the prefactor can be
approximated by $-(k-l)^2 / 2(k+l) \sim - d^2/ (2l_B^2)$ and,
therefore, the leading decaying behavior is essentially the same. So
we will continue to work in the LLL but expect that the leading
scaling behavior is the same as in the 1LL.

In the many-body case, we write the tunneling operator as the sum of
the single-particle operators,
\begin{equation}
{\cal T} = \sum_i V_{\rm tunnel}(\theta_i) = V_t \sum_i \delta(\theta_i).
\end{equation}
We are now ready to calculate the tunneling amplitudes $\Gamma^{e/4} =
\langle \Psi_{\rm MR} \vert {\cal T} \vert \Psi_{\rm MR}^{e/4}
\rangle$ and $\Gamma^{e/2} = \langle \Psi_{\rm MR} \vert {\cal T}
\vert \Psi_{\rm MR}^{e/2} \rangle$ for $e/4$ and $e/2$ quasiholes,
respectively.  For convenience, we will set $V_t = 1$ as the unit of
the tunneling amplitudes in the following text and figures. As
explained in Ref.~\onlinecite{shopen05}, the matrix elements consist
of contributions from the respective Slater-determinant components
$\vert l_1,...,l_N \rangle \in \Psi_{\rm MR}$ and $\vert k_1,...,k_N
\rangle \in \Psi_{\rm MR}^{e/4}$ or $\Psi_{\rm MR}^{e/2}$. Non-zero
contributions enters only when $\vert l_1,...,l_N \rangle$ and $\vert
k_1,...,k_N \rangle$ are identical except for a single pair
$\tilde{l}$ and $\tilde{k}$ with angular momentum difference
$\tilde{k} - \tilde{l} = N/2$ or $N$ for the quasihole with charge
$e/4$ or $e/2$.

For clarity, we also include a pedagogical illustration of the
procedure for calculating the tunneling matrix elements in the
smallest possible system of four electrons in
Appendix~\ref{fewparticle}.

\section{Results}
\label{sec:results}

\subsection{Short-range interaction}
\label{subsec:shortrange}

\begin{figure}
  \includegraphics[width=7cm]{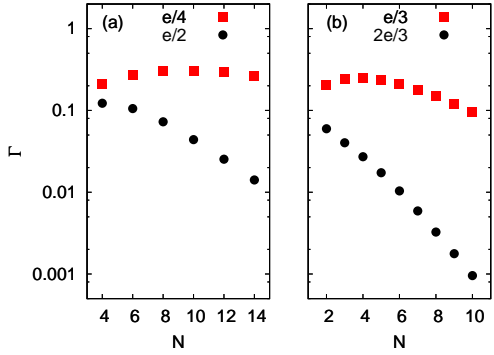}
  \caption{\label{fig:trend}(Color online) tunneling amplitude as
    a function of number of electrons for (a) $e/4$ and $e/2$
    quasiholes in the Moore-Read state at half filling and (b) $e/3$
    and $2e/3$ quasiholes in the Laughlin state at $\nu = 1/3$. }
\end{figure}

Systems of up to six electrons can be worked out pedagogically using
Mathematica as illustrated in Appendix~\ref{fewparticle}.  For larger
systems, we obtain the exact Moore-Read and quasihole wavefunctions by
the exact diagonalization of the three-body Hamiltonian
[Eq.~(\ref{eqn:threebody})] using the Lanczos algorithm. The tunneling
amplitudes are then evaluated as explained in Sec.~\ref{sec:model}.
Figure~\ref{fig:trend}(a) plots the tunneling amplitudes for the $e/4$
and $e/2$ quasiholes in the Moore-Read state as a function of electron
number.  The result for the $e/4$ quasihole shows a weak increase for
$N \le 10$ followed by a decrease for $N > 10$. On the other hand, the
result for the $e/2$ quasihole shows a monotonic decrease as the
number of electrons increases up to 14. In the largest system, the
ratio of the two tunneling matrix elements is slightly less than 20.
For comparison, we also plot the tunneling amplitudes $\Gamma^{e/3}$
and $\Gamma^{2e/3}$ for the $e/3$ and $2e/3$ quasiholes in a Laughlin
state at $\nu = 1/3$ in Fig.~\ref{fig:trend}(b). We also observe a
bump in the tunneling amplitude $\Gamma^{e/3}$ for charge $e/3$,
followed by a monotonic decrease. We thus expect the $\Gamma^{e/4}$
would eventually also show a monotonic decrease for large enough
systems. $\Gamma^{2e/3}$ for charge $2e/3$ shows a much faster
decrease, consistent with its larger charge, and thus a larger
momentum transfer for the same tunneling distance.

\begin{figure}
\includegraphics[width=7cm]{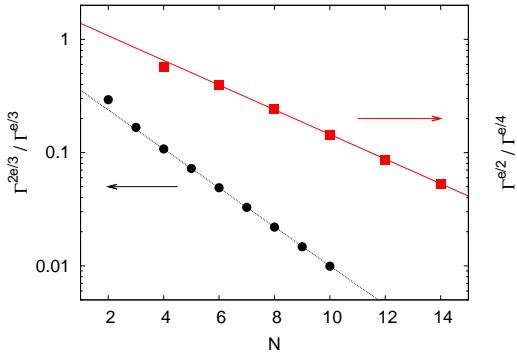}
\caption{\label{fig:ratio}(Color online) The ratio of tunneling
  matrix elements for $e/2$ quasiholes to $e/4$ quasiholes in the
  Moore-Read state at half filling, and for $2e/3$ quasiholes to $e/3$
  quasiholes in the Laughlin state at $1/3$ filling as a function of
  number of electrons. }
\end{figure}

Due to the finite-size bumps in $\Gamma^{e/4}$ and $\Gamma^{e/3}$ for
charge $e/4$ and $e/3$, it is difficult to extract the asymptotic
behavior in the tunneling amplitudes for these
quasiparticles. However, we may expect that such finite-size
corrections also exist in the tunneling amplitude for charge $e/2$ and
$2e/3$ so we can extract the asymptotic behavior in their
ratios. Fortunately, this is indeed the case. We plot $\Gamma^{e/2} /
\Gamma^{e/4}$ and $\Gamma^{2e/3} / \Gamma^{e/3}$ in
Fig.~\ref{fig:ratio}. We find the ratios can be fitted very well by
exponentially decaying trends for almost all finite system sizes.  The
fitting results are
\begin{eqnarray}
\label{eqn:ratio}
\Gamma^{e/2} / \Gamma^{e/4} &\simeq& 1.78 e^{-0.25N}, \\
\Gamma^{2e/3} / \Gamma^{e/3} &\simeq& 0.53 e^{-0.40N}.
\end{eqnarray}
As will be discussed later, the exponents are related to the charge of
the quasiholes and, to a lesser extent, to corrections due to sample
geometry, perhaps also to the influence of the neutral component of
the charge-e/4 quasiparticles.  Quanlitatively, the constant in the
exponent of the ratio $\Gamma^{e/2} / \Gamma^{e/4}$ is found to be
smaller than that for $\Gamma^{2e/3} / \Gamma^{e/3}$, consistent with
the smaller charge and thus smaller charge difference in the
half-filled case.

One may question that the tunneling amplitude for a quasihole from the
disk center to the disk edge may be different from that for edge to
edge, as in the realistic experimental situations. In particular, the
former can contain a geometric factor, which can be corrected by
mapping the disk to a annulus (or a ribbon) by inserting a large
number of quasiholes at the disk center, from which electrons are
repelled (see Appendix~\ref{mapping} for technical details). Inserting
$n$ quasiholes to the center of a disk of $N$ electrons in the
Moore-Read state, we can write the new wavefunction as
\begin{equation}
\Psi_{\rm MR}^{ne/2} = \left ( \prod_{i=1}^N z_i^n \right )\Psi_{\rm MR},
\end{equation}
so that each component Slater determinant gets shifted into a new one
to be normalized. The first $n$ orbitals from the center are now
completely empty and the electrons are occupying orbitals from $n$ to
$n + 2N - 3$. This transformation, of course, also changes the
tunneling distance to
\begin{equation}
\label{eq:distance}
d(n,N) / l_B = \sqrt{2(n + 2N - 2)}-\sqrt{2n}.
\end{equation}
So we can plot data using $d(n,N)$, rather than $n$. Similarly, we can
make the same transformation for the Moore-Read state with either an
additional charge $e/4$ excitation or an additional charge $e/2$
excitation at the inner edge defined by the inserted $n$ quasiholes.
Thus, we can calculate the tunneling amplitudes under the mapping from
disk to annulus.

\begin{figure}
\includegraphics[width=7cm]{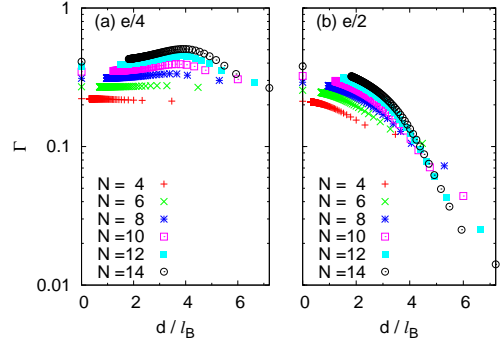}
\caption{\label{fig:tvd}(Color online) (a) The tunneling amplitude
  $\Gamma^{e/4}$ and (b) the tunneling amplitudes $\Gamma^{e/2}$ as
  functions of edge-to-edge distance $d(n, N)$. Data is shown up to $n
  = 100$ quasiholes and $N = 14$ electrons.  The data points at $d =
  0$ or $n \rightarrow \infty$ limit are exact results as explained in
  Appendix~\ref{mapping}.}
\end{figure}

\begin{figure}
\includegraphics[width=7cm]{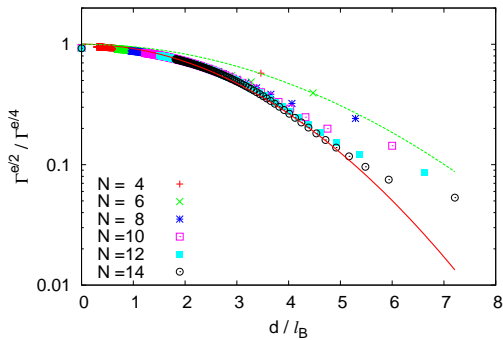}
\caption{\label{fig:rot}(Color online) The ratio of tunneling
  amplitudes $\Gamma^{e/2} / \Gamma^{e/4}$ as a function of
  edge-to-edge distance $d(n, N)$. We also plot Eq.~(\ref{eq:scaling})
  as the solid line to guide the eye. The dashed line is the
  theoretical estimate based on the charge component only
  [Eq.~(\ref{eqn:chargeonly})]. }
\end{figure}

In Fig.~\ref{fig:tvd}, we show the tunneling amplitudes $\Gamma^{e/4}$
and $\Gamma^{e/2}$ for up to $n = 100$ quasiholes. We plot them as
functions of tunneling distance $d$, which decreases as $n$ increases.
It is interesting to note that finite-size effects diminish beyond $d
> 6 l_B$ for charge $e/4$ and $d > 5l_B$ for charge $e/2$. 
For comparison, we plot the ratio $\Gamma^{e/2} /
\Gamma^{e/4}$ as a function of $d$ in Fig.~\ref{fig:rot}. We find
that, when we insert more than one quasihole, the ratio of the tunneling
amplitudes falls onto a single curve, regardless of the system
size $N$ and the number of quasiholes $n$. The curve can be fit roughly to
\begin{equation}
\label{eq:scaling}
{\Gamma^{e/2} (d) \over \Gamma^{e/4} (d)} \sim e^{-0.083 (d/l_B)^2}.
\end{equation}
We point out that a few points in Fig.~\ref{fig:rot} can been seen
deviated from this behavior. They correspond to the largest $d$ for a
given $N$, meaning that there is no quasihole in the bulk, thus
corrspond to the bulk-to-edge instead of the edge-to-edge tunneling. 

It is worth pointing out that such behavior is not completely
unexpected; in fact, it reflects the asymptotic behavior of the
single-particle tunneling matrix and the corresponding charge of the
quasiparticles. To see this, we note that for a charge $q$ quasihole to
tunnel a distance of $d$, one electron (in each Slater determinant)
must hop by a distance of $qd/e$ for the exact momentum
transfer. According to the asymptotic behavior in
Eq.~(\ref{eq:asymptotic}), we expect
\begin{equation}
  \Gamma^{q} \sim e^{-(qd/2el_B)^2}.
\end{equation}
Therefore, we expect
\begin{equation}
\label{eqn:chargeonly}
{\Gamma^{e/2} \over \Gamma^{e/4}}
\sim e^{-[(d/2)^2 - (d/4)^2] / (2 l_B)^2} = e^{-0.047 (d/l_B)^2},
\end{equation}
which we also include in Fig.~\ref{fig:rot} for comparison. 

We thus find that both variational wavefunction calculation and
qualitative analysis suggest that the tunneling amplitude of the $e/2$
quasiparticles is smaller than that of the $e/4$ quasiparticles by a
Gaussian factor in edge-to-edge distance $d$, which is the main
results of this paper. There is, however, a quantitatively discrepancy
in the length scale associated with the Gaussian dependence between
Eqs. (\ref{eqn:chargeonly}) and (\ref{eq:scaling}). This indicates
that the Gaussian factor in single-electron tunneling matrix element
only partially accounts for the Gaussian dependence; the remaining
decaying factor thus must be of many-body origin, whose nature is not
clear at present and warrants further study.

\subsection{Long-range interaction}
\label{subsec:longrange}

So far, we have discussed the tunneling amplitudes using the
variational wavefunctions, which are exact ground states of the
three-body Hamiltonian. These wavefunctions are unique, but in general
not the exact ground states of any generic Hamiltonian one may
encounter in a realistic sample. In reality, long-range Coulomb
interaction is overwhelming, although Landau level mixing can generate
effective three-body interaction.~\cite{bishara09b} In this
subsection, we explore the quasihole tunneling in the presence of
long-range Coulomb interaction. The central questions are the
following. First, how can we generate both non-Abelian and Abelian
quasiholes in practice? Remember now we do not have the variational
Moore-Read state as the exact ground state, so the variational
quasihole states are also less meaningful. We attempt to generate and
localize quasiholes with a single-body impurity potential; then how
close are the corresponding wavefunctions to the variational ones?
Second, suppose we have well-defined quasihole wavefunctions, are the
results on the tunneling amplitude obtained in the short-range
three-body interaction case robust in the presence of long-range
Coulomb interaction?

For a smooth interpolation between the short- and long-range cases, we
introduce a mixed Hamitonian
\begin{equation}
\label{eqn:mixed}
H_{\lambda} = (1-\lambda) H_C + \lambda H_{3B},
\end{equation}
as explained in our earlier works.~\cite{wan06,wan08} Here, the
dimensionless $\lambda$ interpolates smoothly between the limiting
cases of the three-body Hamiltonian $H_{3B}$ ($\lambda=1$) and a
two-body Coulomb Hamiltonian $H_C$ ($\lambda=0$). $H_c$ also includes
a background confining potential arising from neutralizing background
charge distributed uniformly on a parallel disk of radius $R =
\sqrt{4N}$, located at a distance $D$ above the 2DEG.  Using the
symmetric gauge, we can write down the Hamiltonian for electrons in
the 1LL as
\begin{equation}
\label{eqn:chamiltonian} H_{\rm C} = {1\over 2}\sum_{mnl}V_{mn}^l
c_{m+l}^\dagger c_n^\dagger c_{n+l}c_m +\sum_m U_mc_m^\dagger c_m,
\end{equation}
where $c_m^\dagger$ is the electron creation operator for the first
excited Landau level (1LL) single electron state with angular momentum
$m$. $V_{mn}^l$'s are the corresponding matrix elements of Coulomb
interaction for the symmetric gauge, and $U_m$'s the corresponding
matrix elements of the confining potential. We choose $D = 0.6$ so the
ground state can be well described by the Moore-Read state.

To be experimentally relevant, we also want to generate the quasihole
states by a generic impurity potential, rather than by the artificial
interaction [Eq.~(\ref{eqn:repulsion})] we used above to generate the
unique $e/4$ quasihole state in the three-body case. We consider a
Gaussian impurity potential,~\cite{hu08a}
\begin{equation}
\label{eqn:tippotential}
H_{imp} (W,s) = W \sum_m  e^{- m^2 / 2 s^2} c_m^{\dagger} c_m,
\end{equation}
which will trap at the disk center an $e/4$ or $e/2$ quasihole
depending on its strength.~\cite{wan08} Here, $s$ characterizes the
range of the potential. Note $H_{imp} = W c_0^{\dagger} c_0$ is the
short-range limit ($s \rightarrow 0$) of the Gaussian potential in
Eq.~(\ref{eqn:tippotential}). $W$ is always expressed in units of
$e^2/(\epsilon l_B)$.

\begin{figure}
\includegraphics[width=7cm]{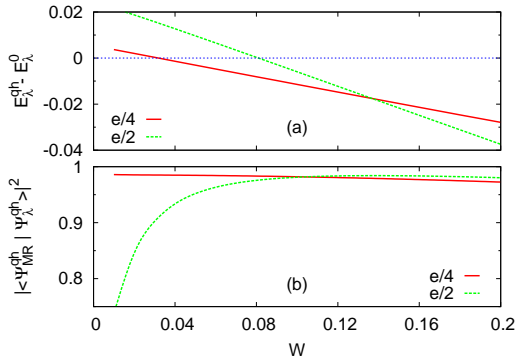}
\caption{\label{fig:overlap} (Color online) (a) Energies of the $e/4$ and $e/2$
  quasihole states $E^{qh}_{\lambda}$, measured from the corresponding
  ground state in the momentum $M_0$ subspace $E^0_{\lambda}$, as a
  function of the strength $W$ of the Gaussian trapping potential at
  the disk center with $s = 2.0 l_B$ for the $\nu = 5/2$ state with a
  mixed Hamiltonian $H_{\lambda}$ ($\lambda = 0.5$). The $e/4$
  quasihole state is energetically favorable for $0.032 < W <
  0.137$. (b) Overlaps of the $e/4$ and $e/2$ quasihole states $\vert
  \Psi_{\lambda}^{qh} \rangle$ for the mixed Hamiltonian with the
  corresponding variational states $\vert \Psi_{MR}^{qh} \rangle$
  [Eqs.~(\ref{eqn:eover4}) and (\ref{eqn:eover2})].}
\end{figure}

Earlier studies~\cite{hu08a,wan08} have identified $s = 2.0$ as a
suitable width for the Gaussian trapping potential, which is of
roughly the radial size of a quasihole. So we use this value
exclusively in the following discussion. One expects that for small
$W$, the system remains in the Moore-Read phase without any quasihole
excitation in the bulk; for later reference, we use $E_{\lambda}^0$ to
denote the ground state energy in the momentum subspace of $M = M_0 =
N(2N-3)/2$. As $W$ increases, the impurity potential first tends to
attract a charge-$e/4$ quasihole, the smallest charge excitation, at
the disk center. This would be reflected in the sudden angular
momentum change from $M_0$ to $M_0 + N/2$ of the global ground state,
which is also characterized by a depletion of $1/4$ of an electron in
the electron occupation number at orbitals with small momentum. We use
$E_{\lambda}^{e/4}$ to denote the ground state energy in the subspace
of $M = M_0 + N/2$. When $W$ is increased further, one can trap a
charge-$e/2$ quasihole at the center, with ground state having the
total angular momentum of $M_0 + N$; in this momentum subspace, we use
$E_{\lambda}^{e/2}$ to denote the ground state energy. We illustrate
this scenario for the case for $\lambda = 0.5$ in
Fig.~\ref{fig:overlap}(a), in which we plot the energies of the $e/4$
and $e/2$ quasihole states $E_{\lambda}^{e/4}$ and
$E_{\lambda}^{e/2}$, measured from the corresponding
$E_{\lambda}^0$. More precisely, the $e/4$ quasihole state is
energetically favorable for $0.032 < W < 0.137$. At $W < 0.032$, we
find $E_{\lambda}^0 < E_{\lambda}^{e/4} < E_{\lambda}^{e/2}$, while at
$W > 0.137$, we find $E_{\lambda}^0 > E_{\lambda}^{e/4} >
E_{\lambda}^{e/2}$.

To understand how good these wavefunctions are, we plot, in
Fig.~\ref{fig:overlap}(b), the overlap of the $e/4$ quasihole state
$\vert \Psi_{\lambda}^{e/4} \rangle$ with the corresponding
variational state $\vert \Psi_{MR}^{e/4} \rangle$
[Eq.~(\ref{eqn:eover4})], as well as the overlap of the $e/2$
quasihole state $\vert \Psi_{\lambda}^{e/2} \rangle$ with the
corresponding variational state $\vert \Psi_{MR}^{e/2} \rangle$
[Eq.~(\ref{eqn:eover2})]. We find that for intermediate $W$ the
overlaps are larger than 97\%.  At small $W$, $\vert
\Psi_{\lambda}^{e/2} \rangle$ does not agree with $\vert
\Psi_{MR}^{e/2} \rangle$ well, but they have very large overlap when
the charge-$e/2$ quasihole state energy $E_{\lambda}^{e/2}$ is lower
than the correponding energy $E_{\lambda}^0$ of the Moore-Read--like
state. On the other hand, it is a little surprising to see the
excellent agreement between $\vert \Psi_{\lambda}^{e/4} \rangle$ and
$\vert \Psi_{MR}^{e/4} \rangle$, as they are generated by different
Hamiltonian [Eqs.~(\ref{eqn:mixed}) and (\ref{eqn:threebody})] with
different trapping potential [Eqs.~(\ref{eqn:tippotential}) and
  (\ref{eqn:repulsion})] respectively. But we note that the smooth
Gaussian trapping potential does favor the charge-$e/4$ quasihole
state, which has no simultaneous occupation of the lowest two
orbitals.

\begin{figure}
\includegraphics[width=7cm]{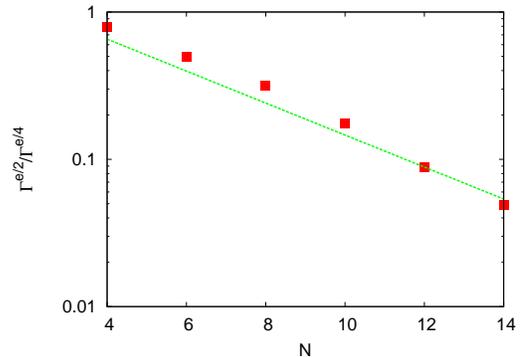}
\caption{\label{fig:gaussratio} (Color online) The ratio of the tunneling amplitudes
  for a mixed Hamiltonian as a function of number of electrons. The
  mix parameter $\lambda = 0.5$ and the width and strength of Gaussian
  potential are $W = 0.1$ and $s = 2.0$, respectively. The background
  potential is located at the distance $D = 0.6l_B$. The dotted line
  is the exponentially decaying trend line [Eq.~(\ref{eqn:ratio})] for
  the pure three-body case as shown in Fig.~\ref{fig:ratio}. }
\end{figure}

Therefore, we expect that with a moderate mixture of the long-range
Coulomb interaction, the results on the tunneling amplitudes are
rather robust. In particular, we choose $W = 0.1$, at which we have
$E_{\lambda}^0 > E_{\lambda}^{e/2} > E_{\lambda}^{e/4}$ and at which
both $\vert \langle \Psi_{MR}^{e/4} \vert \Psi_{\lambda}^{e/4} \rangle
\vert^2$ and $\vert \langle \Psi_{MR}^{e/2} \vert \Psi_{\lambda}^{e/2}
\rangle \vert^2$ are very close to 1. For example, we plot the ratio
of tunneling amplitudes $\Gamma^{e/2} / \Gamma^{e/4}$ as a function of
the number of electrons for $\lambda = 0.5$ in
Fig.~\ref{fig:gaussratio}. The data points are in good agreement with
the trend [Eq.~(\ref{eqn:ratio})] obtained earlier for the pure
three-body case.

\begin{figure}
\includegraphics[width=7cm]{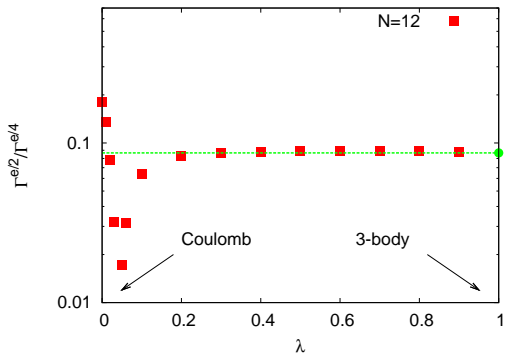}
\caption{\label{fig:coulomb} (Color online) The ratio of tunneling amplitudes
  $\Gamma^{e/2} / \Gamma^{e/4}$ as a function of the mixing parameter
  between the three-body interaction and Coulomb interaction in a
  12-electron system at half-filling in case of $D=0.6l_B$. The width
  and strength of Gaussian potential are $W = 0.1$ and $s = 2.0$
  respectively. The green dot at $\lambda = 1$ is the value for the
  short-range three-body interaction case obtained in
  Sec.~\ref{subsec:shortrange}.  }
\end{figure}

So far, we have shown a case where the presence of the long-range
interaction has very weak effects on the results of tunneling
amplitudes. However, in general, one can expect such an agreement
becomes worse as one move farther away from the pure repulsive
three-body interaction in the parameter space. To present a more
quantitative picture, we plot in Fig.~\ref{fig:coulomb} the ratio of
tunneling amplitudes $\Gamma^{e/2} / \Gamma^{e/4}$ (without inserting
quasiholes at the center, i.e. $n = 0$) as a function of $\lambda$ for
the 12-electron system with the mixed Hamiltonian $H_{\lambda}$ and a
Gaussian trapping potential ($W=2.0$, $s=1.0$). The ratio remains as a
constant from $\lambda = 1$ down to 0.2, before it fluctuates
significantly; the fluctuation is believed to be related to the
stripe-like phase near the pure Coulomb case in finite systems, as
also revealed in earlier work.~\cite{wan08} We point out that recent
numerical work suggests that the spin-polarized Coulomb ground state
at $\nu=5/2$ is adiabatically connected with the Moore-Read wave
function for systems on the surface of a sphere,~\cite{storni09} so
the large deviation may well be a finite-size artifact.

Varying parameters, such as $W$, $s$, and $D$, can also lead to larger
deviation from the pure three-body case, although we find in generic
cases $\Gamma^{e/2} / \Gamma^{e/4}$ remains small.  We remind the
reader that the Moore-Read phase is extremely fragile. Therefore, we
have rather strong constraints on parameters when both the Moore-Read--like
ground state and the quasihole states should subsequently be good
description of the ground states as the impurity potential strength
increases.  For example, the window of $D$ for the ground state at $W
= 0$ to be of Moore-Read nature in the pure Coulomb case is very
narrow ($0.51 < D/l_B < 0.76$ for 12 electrons in 22
orbits~\cite{wan06}); in this range, the effect of the background
potential parameter $D$ on the ratio of tunneling amplitudes is
negligible (less than 1\% variation). Therefore, we conclude that the
small ratio of $\Gamma^{e/2} / \Gamma^{e/4}$ is robust in the presence
of the long-range interaction as long as the system remains in the
Moore-Read phase.

\section{Discussion}
\label{sec:discussion}

In this work we use a simple microscopic model to study quasiparticle
tunneling between two fractional quantum Hall edges.  We find the
tunneling amplitude ratio of quasiparticles with different charges
decays with a Gaussian tail as edge-to-edge distance increases.  The
characteristic length scale associated with this dependence can be
partially accounted for by the difference in the charges of the
corresponding quasiparticles.  More specifically, we find the
tunneling amplitude for a charge $e/4$ quasiparticle is significantly
larger than a charge $e/2$ quasiparticle in the Moore-Read quantum
Hall state, which may describe the observed fractional quantum Hall
effect at the filling factor $\nu=5/2$. This result was anticipated in
Ref. \onlinecite{bishara09}, in which the authors outlined a
microscopic calculations that is similar to the discussion in
Sec.~\ref{subsec:shortrange} (see their Appendix B).

It is worth emphasizing that what we have calculated here are the {\em
  bare} tunneling amplitudes. Under renormalization group (RG)
transformations, both amplitudes will {\em grow} as one goes to lower
energy/temperature, as they are both relevant couplings in the RG
sense. The ratio between them, $\Gamma^{e/2} / \Gamma^{e/4}$, will
{\em decrease} under RG, because $\Gamma^{e/4}$ is {\em more} relevant
than $\Gamma^{e/2}$, which renders $\Gamma^{e/2}$ even less important
than $\Gamma^{e/4}$ at low temperatures. This is clearly consistent
with tunneling experiments involving a {\em single} point
contact,\cite{dolevnature,RaduScience} where only signatures of $e/4$
tunneling is seen.

\begin{figure}
\includegraphics[width=7cm]{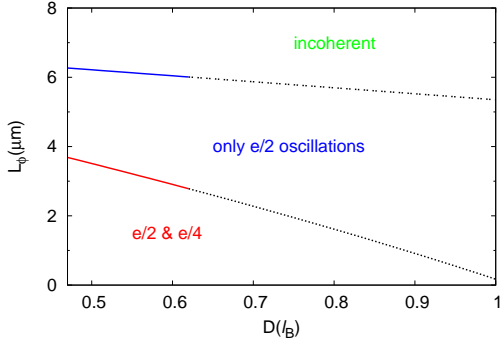}
\caption{\label{fig:coherence} (Color online) Decoherence length
  $L_{\phi}$ as a function of $D$ for both e/4 (upper line) and e/2
  (lower line) quasipaticles in the Moore-Read Pfaffian state. We
  choose a temperature $T = 25$ mK to allow a direct comparison with
  experiment.~\cite{willett08} The broken lines above $D = 0.62 l_B$
  are obtained by extrapolation, as the Moore-Read--like ground state
  is no longer stable in a system of 12 electrons in 26 orbitals. We
  note a stripe phase may emerge below $D = l_B$.~\cite{wan08} }
\end{figure}

However, the importance of the two kinds of quasiparticles can be
reversed in interferometry experiments that look for signatures from
interference between {\em two} point contacts.  This is because the
interference signal depends not only on the quasiparticle tunneling
amplitudes, but also their coherence lengths when propagating along
the edge of fractional quantum Hall samples. Recently, Bishara and
Nayak~\cite{bishara08} found that in a double point-contact
interferometer, the oscillating part of the current for charge $q$
quasiparticles can be written as
\begin{equation}
I_{12}^{(q)} \propto \gamma \vert \Gamma_1^{(q)} \vert
\vert \Gamma_2^{(q)} \vert e^{-\vert x_{12} \vert / L^{(q)}_{\phi}}
\cos \left ( {{2 \pi q \Phi} \over {e \Phi_0}} + \delta^{(q)} + \alpha \right ),
\end{equation}
where $\Gamma_{1,2}^{(q)}$ are the charge $q$ quasiparticle tunneling
amplitudes at the two quantum point contacts 1 and 2 with a distance
of $x_{12}$. $\gamma$ is a suppression factor resulting from the
possible non-Abelian statistics of the quasiparticles. For $q = e/2$,
we have $s = 1$, while for $q = e/4$, $\gamma = \pm 1/\sqrt{2}$ (or 0)
when we have even (or odd) number of $e/4$ quasiparticles in bulk. The
sign depends on whether the even number of $e/4$ quasiparticles fuse
into the identity channel ($+$) or the fermionic channel ($-$). $\Phi$
is the flux enclosed in the interference loop and $\Phi_0 = hc / e$ is
the magnetic flux quantum.  The phase $\delta^{(q)}$ is the
statistical phase due to the existence of bulk quasiparticles inside
the loop and $\alpha$ the phase $\arg(\Gamma_1 \Gamma_2^*)$.  At a
finite temperature $T$, the decoherence length $L^{(q)}_\phi$ for the
quasiparticle in the Moore-Read state is\cite{bishara08}
\begin{equation}
L^{(q)}_{\phi} = {1 \over {2 \pi T}} \left ({g_c^{(q)} \over v_c}
+ {g_n^{(q)} \over v_n} \right )^{-1},
\end{equation}
where $v_{c,n}$ are the charge and neutral edge mode velocities and
$g_{c,n}^{(q)}$ the charge and neutral sector scaling exponents for
charge $q$ quasiparticles, respectively. Earlier studies by the
authors~\cite{wan06,wan08} found that the neutral velocity can be
significantly smaller (by a factor of 10) than the charge velocity,
leading to a shorter coherence length $L_{\phi}^{(1/4)}$ for charge
$e/4$ quasiparticles (less than $1/3$ of $L_{\phi}^{(1/2)}$ for charge
$e/2$ quasiparticles in the Moore-Read case, as $L_{\phi}^{(1/2)}$
depends on $v_c$ only because it is Abelian and
$g_n^{(1/2)}=0$). Finite-size numerical analysis~\cite{hu09} maps out
the dependence of $L_{\phi}^{(1/4)}$ and $L_{\phi}^{(1/2)}$ on the
strength of confining potential parametrized by $D$ for the Moore-Read
state, as summarized in Fig.~\ref{fig:coherence} for $T = 25$ mK used
in the recent experimental study.~\cite{willett08} Depending on the
size of the interference loop and the strength of the confining
potential, the edge transport may exhibit both $e/4$ and $e/2$
quasiparticle interference, $e/2$ quasiparticle interference only, or
no quasiparticle interference. In particular, the observation of the
$e/4$ quasiparticle interference depends sensitively on the length of
the interference loop due to the effect of the confining potential
strength on the neutral mode velocity $v_n$.

We close by stating that by combining the small ratio between $e/4$
and $e/2$ quasiparticle tunneling matrix elements and the fact that
$e/2$ quasiparticle has longer coherence length along the edge, it is
possible to provide a consistent interpretation of the recent
tunneling\cite{dolevnature,RaduScience} and
interference\cite{willett08,willett09} experiments.  Similar
conclusions were reached in a recent comprehensive
analysis~\cite{bishara09} of the interference
experiments.~\cite{willett08,willett09} We would like to caution
though that a complete quantitative understanding of the experiments
is not yet available at this stage due to our incomplete understanding
of the actual ground state, mesoscopic effects, and the possible
oversimplifications of the microscopic model (for example edge
reconstruction~\cite{wan02,yang03,overbosch08} may occur and
complicate the analysis significantly). Nonetheless, we hope that the
quantitative analysis presented here can help solve the puzzle.

\section*{Acknowledgment}

This work was supported by NSF grants No. DMR-0704133 (K.Y.) and
DMR-0606566 (E.H.R.), as well as PCSIRT Project No. IRT0754
(X.W.). X.W. acknowledges the Max Planck Society and the Korea
Ministry of Education, Science and Technology for the joint support of
the Independent Junior Research Group at the Asia Pacific Center for
Theoretical Physics. K.Y. was visiting the Kavli Institute for
Theoretical Physics (KITP) during the completion of this work. The
work at KITP was supported in part by National Science Foundation
grant No. PHY-0551164. 

\appendix
\section{Single-particle tunneling matrix elements in the large distance limit}
\label{larged}

In the disk geometry, the single-particle eigenstates are
\begin{equation}
\vert m \rangle \equiv \phi_m(z) = (2\pi2^m m!)^{-1/2} z^m e^{-|z|^2/4}.
\end{equation}
If we assume a single-particle tunneling potential
\begin{equation}
V_{\rm tunnel} = V_t \delta(\theta),
\end{equation}
the matrix element of $\langle k \vert V_t \vert l
\rangle$, related to the tunneling of an electron from state $\vert
l \rangle$ to state $\vert k \rangle$, is
\begin{equation}
v_p (k,l) \equiv \langle k \vert V_{\rm tunnel} \vert l \rangle = {V_t \over 2 \pi}
{\Gamma \left ( {{k + l} \over 2} + 1 \right ) \over \sqrt{k! l!}}.
\end{equation}

Using beta functions
\begin{equation}
B(x,y)=\frac{\Gamma\left( x\right) \Gamma\left( y\right) }{\Gamma\left( x+y\right) },
\end{equation}
we can rewrite the dimensionless tunneling matrix element as
\begin{equation}
\tilde{v}_p (k,l) \equiv \frac{ 2 \pi v_p (k,l)}{V_t} = \left [ \frac{B\left ( {{k + l} \over 2} + 1 , {{k + l} \over 2} + 1 \right ) }{
B\left ( k+1,l+1 \right)} \right ] ^{1/2}.
\end{equation}
We are interested in the limit of large $l$ and large $k$, where we can use the asymptotic formula of Stirling's approximation
\begin{equation}
B(x,y)\sim \sqrt{2\pi}\frac{x^{x-1/2}y^{y-1/2}}{\left( x+y\right) ^{x+y-1/2}}
\end{equation}
for large $x$ and large $y$. Therefore, we have
\begin{equation}
\tilde{v}_p (k,l) \sim \left [ \frac{ \left( \frac{k+l}{2}+1 \right)^{k+l+1} }
{\left( k+1\right) ^{k+1/2}\left( l+1 \right) ^{l+1/2}} \right ]^{1/2}.
\end{equation}
For convenience, we define
\begin{equation}
S=\frac{k+l}{2} \ {\rm and} \ M=\frac{k-l}{2}.
\end{equation}
If we further take the limit of $S \gg |M|$, we find
\begin{eqnarray}
  \tilde{v}_p (k,l) &\sim& \left( 1+\frac{M}{S+1}\right) ^{ -{{S+M } \over 2} - {1 \over 4} }
  \left( 1-\frac{M}{S+1}\right) ^{ -{{S-M } \over 2} - {1 \over 4}} \nonumber \\
  &\sim& \left [1-\left ({M \over {S + 1}} \right )^2 \right ]^{-{S \over 2} - {1 \over 4}} \left (1-{M \over {S + 1}} \right )^M \nonumber \\
  &\sim& e^{{M^2(S+1/2) \over 2(S+1)^2}} e^{-{M^2 \over {S+1}}}   \nonumber \\
  &\sim&e^{-{M^2 \over 2(S+1)} }.
\end{eqnarray}

\section{Tunneling matrix elements in few-electron systems}
\label{fewparticle}

In this appendix, we first illustrate the calculation of the tunneling
matrix elements in a four-electron system for the
Moore-Read state.  In this case, the normalized Moore-Read
wavefunction can be written as a sum of Slater determinants as
\begin{equation}
\label{eqn:mr}
\Psi_{\rm MR} =  {{
\sqrt{10} \vert 011110 \rangle
- \sqrt{2} \vert 101101 \rangle
+ \vert 110011 \rangle } \over \sqrt{13}},
\end{equation}
where the ket notation denotes a Slater determinant with electrons
occuping the single-particle orbitals labeled by 1. For example,
$\vert 110011 \rangle$ means the normalized antisymmetric wavefunction
of four electrons occupying the orbitals with angular momentum 0,1,5,6
(reading from left to right in the ket).  This can be obtained by
explicitly expanding the Moore-Read state with four electrons (trivial
with the help of Mathematica). The corresponding $e/4$ quasihole
state, similarly, can be written as
\begin{eqnarray}
\Psi_{\rm MR}^{e/4} &=&  {1 \over 5 \sqrt{11}} \left (
\sqrt{3} \vert 1010101 \rangle
- 4 \sqrt{6} \vert 1001110 \rangle \right . \nonumber \\
&+& \left . 8 \sqrt{2} \vert 0110110 \rangle
- 4 \sqrt{3} \vert 0111001 \rangle
\right ),
\end{eqnarray}
and the $e/2$ quasihole state
\begin{eqnarray}
\label{eqn:mr3}
\Psi_{\rm MR}^{e/2} &=&  {1 \over 3 \sqrt{13}} \left (
10 \vert 0011110 \rangle  \right . \nonumber \\
&-& \left . 2 \sqrt{3} \vert 0101101 \rangle
+ \sqrt{5} \vert 0110011 \rangle \right ).
\end{eqnarray}

In the many-body case, we write the tunneling operator as the sum of
the single-particle operators,
\begin{equation}
{\cal T} = V_t \sum_i \delta(\theta_i),
\end{equation}
and calculate the tunneling amplitudes $\Gamma^{e/4} = \langle
\Psi_{\rm MR} \vert {\cal T} \vert \Psi_{\rm MR}^{e/4} \rangle$ and
$\Gamma^{e/2} = \langle \Psi_{\rm MR} \vert {\cal T} \vert \Psi_{\rm
  MR}^{e/2} \rangle$ for $e/4$ and $e/2$ quasiholes, respectively.
The matrix elements consist of contributions from the respective
Slater-determinant components $\vert l_1,...,l_N \rangle \in \Psi_{\rm
  MR}$ and $\vert k_1,...,k_N \rangle \in \Psi_{\rm MR}^{e/4}$ or
$\Psi_{\rm MR}^{e/2}$. There are non-zero
contributions only when the two sets $\{l_1,...,l_N\}$ and
$\{k_1,...,k_N\}$ are identical except for a single pair $\tilde{l}$ and
$\tilde{k}$ with angular momentum difference $\tilde{k} - \tilde{l} =
N/2$ or $N$ for the quasihole with charge $e/4$ or $e/2$. One should
also pay proper attention to fermionic signs.

With some algebra, one obtains, for the four-electron case,
\begin{eqnarray}
\langle \Psi_{\rm MR} \vert {\cal T} \vert \Psi_{\rm MR}^{e/4}
\rangle &=& {1 \over 5 \sqrt{143}} \left [
16 \sqrt{5} v_p(3,5) \nonumber  \right . \nonumber \\
&+& 4 \sqrt{30} v_p(4,6)
+ 8 \sqrt{3} v_p(2,4) \nonumber \nonumber \\
&+& \left . 8 \sqrt{2} v_p(0,2) + 4 \sqrt{6} v_p(1,3) \right ]
\end{eqnarray}
and
\begin{eqnarray}
\label{eqn:mr2}
\langle \Psi_{\rm MR} \vert {\cal T} \vert \Psi_{\rm MR}^{e/2}
\rangle &=& {1 \over 39} \left [10 \sqrt{10} v_p(1,5)
+ 10 \sqrt{2} v_p(0,4) \right . \nonumber \\
&+& \left .
2 \sqrt{30} v_p(2,6) \right ],
\end{eqnarray}
where, as before, we define
\begin{equation}
v_p (k,l) = {V_t \over 2 \pi}
{\Gamma \left ( {{k + l} \over 2} + 1 \right ) \over \sqrt{k! l!}}.
\end{equation}

The numerical values for the two tunneling matrix elements are 0.213
and 0.123, respectively, in units of $V_t$. Therefore, in the smallest
nontrivial system, we find that the tunneling amplitude for $e/4$
quasiholes is roughly twice  as large as that for $e/2$ quasiholes.
The example of the four-electron case illustrates how the tunneling
amplitudes can be computed. The results are, however, not particularly
meaningful as the system size is so small that one cannot really
distinguish bulk from edge.

A similar analysis can be performed for a system of six electrons with
the help of Mathematica. Due to larger Hilbert space, we will not
explicitly write down the decomposition of the ground states and
quasihole states by Slater determinants. Instead, we only point out
that the tunneling matrix elements are given by
\begin{equation}
\langle \Psi_{\rm MR} \vert {\cal T} \vert \Psi_{\rm MR}^{e/4}
\rangle = 0.267
\end{equation}
and
\begin{equation}
\langle \Psi_{\rm MR} \vert {\cal T} \vert \Psi_{\rm MR}^{+e/2}
\rangle = 0.105,
\end{equation}
in units of $V_t$.

\section{Mapping from disk to annulus}
\label{mapping}

Microscopic quantum Hall calculations are commonly based on one of the
following geometries (or topologies): torus, sphere, annulus (or
cylinder), and disk. In a specific calculation, they are chosen either
for convenience, or the need for having different numbers of
edge(s). On the other hand one can also map one geometry to another by
means of, e.g., quasihole insertion. Here, to connect the theoretical
analysis with experiment, we perform a mapping from the disk to the
annulus geometry by inserting a large number of quasiholes at the
center of the disk, effectively creating an inner edge, as the
electron density in the center is suppressed by inserting a small disk
of Laughlin quasihole liquid.

After inserting $n$ charge $e/2$ Laughlin quasiholes to the center of an
$N$-electron Moore-Read state, the ground state can be written as
\begin{equation}
\label{eqn:mrn2}
\Psi_{\rm MR}^{ne/2} = \left ( \prod_{i=1}^N z_i^n \right )\Psi_{\rm MR},
\end{equation}
where the additional factor transforms each Slater determinant into a
new one to be normalized.  Let us use the case of four electrons as in
Appendix~\ref{fewparticle} to illustrate. We note, a Slater
determinant $\vert 011110 \rangle$ with an addition of $n$ Laughlin
quasiholes evolves into another Slater determinant $\vert (0^n)011110
\rangle$, meaning that the $m$-th (in this example, $m = 1$-4)
single-particle orbital is now mapped to the $(m+n)$-th orbital. Due
to the difference in normalization, the latter determinant should be
multiplied by a factor of $F(n; 1,2,3,4)$ with a general form of
\begin{equation}
F(n; m_1, m_2, \cdots, m_N) = 2^{nN/2} \prod_{i=1}^N \sqrt{
{(n+m_i)! \over m_i!}}.
\end{equation}
Therefore, when we express Eq.~(\ref{eqn:mrn2}) explicitly for
Eq.~(\ref{eqn:mr}), we have
\begin{multline}
  \Psi_{\rm MR}^{ne/2} = {\cal N} \left [ F(n; 1,2,3,4)
    \frac{\sqrt{10}}
    {\sqrt{13}} \vert (0^n)011110 \rangle  \right . \\
  - F(n; 0,2,3,5) \frac{\sqrt{2}}{\sqrt{13}}\vert (0^n)101101 \rangle \\
  + \left . F(n; 0,1,4,5) \frac{1}{\sqrt{13}}\vert (0^n)110011 \rangle
  \right ],
\end{multline}
where ${\cal N}$ is a numerical normalization factor.  For $n = 1$ we
thus obtain exactly Eq.~(\ref{eqn:mr3}) as expected.  Interestingly,
in the $n \rightarrow \infty$ (ring) limit, the normalized
wavefunction becomes, aymptotically,
\begin{multline}
  \Psi_{\rm MR}^{ne/2} = C \left [ \sqrt{ \frac{ 1 } { 1! 2! 3! 4!} }
    \frac{\sqrt{10}}{\sqrt{13}}\vert (0^n)011110
    \rangle \right . \\
  - \sqrt{ \frac{ 1 }{ 0! 2! 3! 5!} } \frac{\sqrt{2}}{\sqrt{13}}
  \vert (0^n)101101 \rangle \\
  + \left . \sqrt{ \frac{ 1 }{ 0! 1! 4! 5!} } \frac{1}{\sqrt{13}}\vert
    (0^n)110011 \rangle \right ],
\end{multline}
where the normalization factor $C$ is, explicitly,
\begin{equation}
{1 \over C} = \sqrt {
\frac{ 1 }{  1!  2!  3!  4!}\frac{10}{13}
+\frac{ 1 }{  0!  2!  3!  5!} \frac{2}{13}
+\frac{ 1 }{  0!  1!  4!  5!} \frac{1}{13} }.
\end{equation}
In the limit of $n \gg N$, we have
$v_p(n+m_1,n+m_2) \rightarrow V_t / (2 \pi)$.
Therefore, the tunneling matrix between the states with
$n$ quasiholes and $n+1$ quasiholes becomes
\begin{multline}
\label{eqn:tlim}
\langle \Psi_{\rm MR}^{ne/2} \vert {\cal T} \vert \Psi_{\rm MR}^{ne/2+e/2} \rangle \\
=  \frac{V_tC^2}{2\pi} \left [ \frac{ 1 }{1! 2! 3! 4!}  \frac{10}{13} + \sqrt{ \frac{ 1 }{   1! 2! 3! 4!} } \frac{\sqrt{10}}{\sqrt{13}}
 \sqrt{ \frac{ 1 }{  0! 2! 3! 5!} } \frac{\sqrt{2}}{\sqrt{13}} \right . \\
+ \left . \sqrt{ \frac{ 1 }{  0! 2! 3! 5!} } \frac{\sqrt{2}}{\sqrt{13}}
 \sqrt{ \frac{ 1 }{   1! 2! 3! 4!} } \frac{\sqrt{10}}{\sqrt{13}} \right ].
\end{multline}
The tunneling of $e/4$ quasiholes can be worked out in a
similar fashion and we can obtain generically $\left\langle
  \Psi_{MR}^{ne/2}\mid{\cal T}\mid\Psi_{MR}^{ne/2+e/4}\right\rangle$.


\begin{thebibliography}{99}

\bibitem{willett87}
R.~L. Willett, J.~P. Eisenstein, H. L. Stormer, D. C. Tsui,
A.~C. Gossard, and J.~H. English, Phys. Rev. Lett. {\bf 59}, 1776
(1987).

\bibitem{Panw99a}
P. L. Gammel, D. J. Bishop, J. P. Eisenstein,
J. H. English, A. C. Gossard, R. Ruel, and H. L. Stormer,
Phys. Rev. B {\bf 38}, 10 128 (1988).

\bibitem{Panw99b}
J. P. Eisenstein, R. L. Willett, H. L. Stormer, L. N. Pfeiffer, and K. W. West,
Surf. Sci. {\bf 229}, 31 (1990).

\bibitem{Panw99c}
W. Pan, J.-S. Xia, V. Shvarts, D. E. Adams, H. L. Stormer, D. C. Tsui,
L. N. Pfeiffer, K. W. Baldwin, and K. W. West,
Phys. Rev. Lett. {\bf 83}, 3530 (1999).

\bibitem{Panw99d}
W. Pan, H. L. Stormer, D. C. Tsui, L. N. Pfeiffer,
K. W. Baldwin, and K. W. West,
Solid State Commun. {\bf 119}, 641 (2001).

\bibitem{Panw99e}
J. P. Eisenstein, K. B. Cooper, L. N. Pfeiffer, and
K. W. West, Phys. Rev. Lett. {\bf 88}, 076801 (2002).

\bibitem{Panw99f}
J. B. Miller, I. P. Radu, D. M. Zumbuhl, E. M. Levenson-Falk, M. A. Kastner,
C. M. Marcus, L. N. Pfeiffer, and K. W. West,
Nature Phys. {\bf 3}, 561 (2007).

\bibitem{Panw99g}
H. C. Choi, W. Kang, S. Das Sarma, L. N. Pfeiffer, and K. W. West,
Phys. Rev. B {\bf 77}, 081301(R) (2008).

\bibitem{Panw99h}
W. Pan, J. S. Xia, H. L. Stormer, D. C. Tsui, C. Vicente,
E. D. Adams, N. S. Sullivan,
L. N. Pfeiffer, K. W. Baldwin, and K. W. West,
Phys. Rev. B {\bf 77}, 075307 (2008).

\bibitem{Panw99i}
C. R. Dean, B. A. Piot, P. Hayden, S. Das Sarma, G. Gervais,
L. N. Pfeiffer, and K. W. West, Phys. Rev. Lett. {\bf 100}, 146803 (2008).

\bibitem{kitaev}
A. Kitaev, Ann. Phys. {\bf 303}, 2 (2003).

\bibitem{Preskillbook}
J. Preskill, {\it Introduction to Quantum Computation}, edited by H.-K.
Lo, S. Popescu, and T.~P. Spiller (World Scientific, 1998).

\bibitem{Freedman}
M. H. Freedman, Proc. Natl. Acad. Sci. USA {\bf 95}, 98 (1998).

\bibitem{Nayakrmp}
C. Nayak, S. H. Simon, A. Stern, M. Freedman, and S. Das Sarma,
Rev. Mod. Phys. {\bf 80}, 1083 (2008).

\bibitem{morf98} R. H. Morf, Phys.
Rev. Lett. {\bf 80}, 1505 (1998).

\bibitem{Ed2000} E. H. Rezayi and
F. D. M. Haldane, Phys. Rev. Lett. {\bf 84}, 4685 (2000).

\bibitem{wan06}
X. Wan, K. Yang, and E. H. Rezayi,
Phys. Rev. Lett. {\bf 97}, 256804 (2006).

\bibitem{wan08}
X. Wan, Z.-X. Hu, E. H. Rezayi, and K. Yang,
Phys. Rev. B {\bf 77}, 165316 (2008).

\bibitem{peterson}
M. R. Peterson, Th. Jolicoeur, and S. Das Sarma,
Phys. Rev. B {\bf 78}, 155308 (2008);
M. R. Peterson, Th. Jolicoeur, and S. Das Sarma,
Phys. Rev. Lett. {\bf 101}, 016807 (2008).

\bibitem{moller}
G. M\"oller and S. H. Simon,
Phys. Rev. B {\bf 77}, 075319 (2008).

\bibitem{feiguin}
A.~E.~Feiguin, E.~H. Rezayi, K. Yang, C. Nayak, and S. Das Sarma,
Phys. Rev. B {\bf 79}, 115322 (2009).

\bibitem{storni09}
M. Storni, R. H. Morf, and S. Das Sarma, arXiv:0812.2691.

\bibitem{moore91}
G. Moore and N. Read, Nucl. Phys. B {\bf 360}, 362 (1991).

\bibitem{lee07}
S.-S. Lee, S. Ryu, C. Nayak, and M.~P.~A. Fisher, Phys. Rev.
Lett. 99, 236807 (2007).

\bibitem{levin07}
M. Levin, B.~I. Halperin, and B. Rosenow, Phys. Rev. Lett. {\bf 99},
236806 (2007).

\bibitem{nayak96}
C. Nayak and F. Wilczek, Nucl. Phys. B {\bf 479}, 529 (1996).

\bibitem{wen}
X. G. Wen, Phys. Rev. B {\bf 43}, 11025 (1991); Phys. Rev. Lett.
{\bf 64}, 2206 (1990); Phys. Rev. B {\bf 44}, 5708 (1991).

\bibitem{kane92}
C. L. Kane and M. P. A. Fisher, Phys. Rev. B {\bf 46}, 15233
(1992); K. Moon, H. Yi, C. L. Kane, S. M. Girvin, and M.~P.~A.
Fisher, Phys. Rev. Lett. {\bf 71}, 4381 (1993); C. L. Kane and M.
P. A. Fisher, Phys. Rev. Lett. {\bf 72}, 724 (1994).

\bibitem{chamonwen}
C. de C. Chamon and X. G. Wen, Phys. Rev. Lett. {\bf 70}, 2605
(1993).

\bibitem{pfendley}
P. Fendley, M. P. A. Fisher, and C. Nayak, Phys. Rev. B {\bf 75},
045317 (2007).

\bibitem{afeiguin}
A. E. Feiguin, P. Fendley, M. P. A. Fisher, and C. Nayak,
Phys. Rev. Lett {\bf 101}, 236801 (2008).

\bibitem{sdas}
S. Das, S. Rao, and D. Sen, Europhys. Lett. {\bf 86}, 37010 (2009).

\bibitem{dolevnature}
M. Dolev, M. Heiblum, V. Umansky, A. Stern, and  D. Mahalu, Nature
{\bf 452}, 829 (2008).

\bibitem{RaduScience}
I. Radu, J. B. Miller, C. M. Marcus, M. A. Kastner, L. N.
Pfeiffer, and K. W. West, Science {\bf 320}, 899 (2008).

\bibitem{chamon97}
C. de C. Chamon, D. E. Freed, S. A. Kivelson, S. L. Sondhi, and X.
G. Wen, Phys. Rev. B {\bf 55}, 2331 (1997).

\bibitem{fradkin98}
E. Fradkin, C. Nayak, A. Tsvelik, and F. Wilczek, Nucl. Phys. B
{\bf 516}, 704 (1998).

\bibitem{dassarma05}
S. Das Sarma, M. Freedman and C. Nayak, Phys. Rev. Lett. {\bf 94},
166802 (2005).

\bibitem{stern06}
 A. Stern and B. I. Halperin, Phys. Rev. Lett.
{\bf 96}, 016802 (2006).

\bibitem{bonderson06}
P. Bonderson, A. Kitaev, and K. Shtengel, Phys. Rev. Lett. {\bf
96}, 016803 (2006).

\bibitem{rosenow}
B. Rosenow, B. I. Halperin, S. H. Simon, and A.
Stern, Phys. Rev. Lett. {\bf 100}, 226803 (2008).

\bibitem{bishara08}
W. Bishara and C. Nayak,
Phys. Rev. B {\bf 77}, 165302 (2008).

\bibitem{bonderson06b}
P. Bonderson, K. Shtengel, and J. K. Slingerland,
Phys. Rev. Lett. {\bf 97}, 016401 (2006).

\bibitem{fidkowski07}
L. Fidkowski, arXiv:0704.3291.

\bibitem{bonderson08}
P. Bonderson, K. Shtengel, and J. K. Slingerland,
Ann. Phys. {\bf 323}, 2709 (2008).

\bibitem{ardonne08}
E. Ardonne and E.-A. Kim,
J. Stat. Mech. L04001 (2008).

\bibitem{yji03}
Y. Ji, Y. Chung, D. Sprinzak, M. Heiblum, D. Mahalu, and H.
Shtrikman, Nature {\bf 422}, 415 (2003).

\bibitem{yzhang09}
Y. Zhang, D. T. McClure, E. M. Levenson-Falk, C. M. Marcus, L. N.
Pfeiffer, and K. W. West, Phys. Rev. B {\bf 79}, 241304(R) (2009).

\bibitem{camino05}
F. E. Camino, W. Zhou, and V. J. Goldman, Phys. Rev. B {\bf 72},
075342 (2005).

\bibitem{camino07}
F. E. Camino, W. Zhou, and V. J. Goldman, Phys. Rev. Lett. {\bf
98}, 076805 (2007).

\bibitem{willett08}
R. L. Willett, L. N. Pfeiffer, and K. W. West,
PNAS {\bf 106}, 8853 (2009).

\bibitem{willett09}
R. L. Willett, L. N. Pfeiffer, and
K. W. West, unpublished.

\bibitem{bishara09}
W. Bishara, P. Bonderson, C. Nayak, K. Shtengel,
and J. K. Slingerland, Phys. Rev. B {\bf 80}, 155303 (2009).

\bibitem{shopen05}
E. Shopen, Y. Gefen, and Y. Meir,
Phys. Rev. Lett. {\bf 95}, 136803 (2005).

\bibitem{bishara09b}
W. Bishara and C. Nayak, arXiv:0906.2516.

\bibitem{hu08a}
Z.-X. Hu, X. Wan, and P. Schmitteckert,
Phys. Rev. B {\bf 77}, 075331 (2008).

\bibitem{hu09}
Z.-X. Hu, E. H. Rezayi, X. Wan, K. Yang, arXiv:0908.3563.

\bibitem{wan02}
X. Wan, K. Yang, and E. H. Rezayi,
Phys. Rev. Lett. {\bf 88}, 056802 (2002);
X. Wan, E. H. Rezayi, and K. Yang,
Phys. Rev. B {\bf 68}, 125307 (2003).

\bibitem{yang03}
K. Yang,
Phys. Rev. Lett. {\bf 91}, 036802 (2003).

\bibitem{overbosch08}
B. J. Overbosch and X.-G. Wen,
arXiv:0804.2087.

\end{thebibliography}
\end{document}